\documentclass[12pt]{article}

\usepackage{fullpage}
\usepackage{amsmath, amsthm, amssymb}
\usepackage{verbatim}
\usepackage[T1]{fontenc}
\usepackage[tt=false,type1=true]{libertine}
\usepackage[varqu]{zi4}
\usepackage[libertine]{newtxmath}
\usepackage[sort]{natbib}
\usepackage{xcolor}
\usepackage[pagebackref=true,colorlinks]{hyperref}
\hypersetup{linkcolor=blue,filecolor=blue,citecolor=blue,urlcolor=blue}
\usepackage{graphicx}
\usepackage{authblk}

\usepackage{color}

\usepackage{mathtools}
\usepackage{xspace}
\usepackage{xurl}
\usepackage{comment}

\setcitestyle{authoryear}

\newtheorem{theorem}{Theorem}[section]

\theoremstyle{definition}
\newtheorem{definition}[theorem]{Definition}

\newcommand{\cA}{\mathcal{A}}
\newcommand{\cI}{\mathcal{I}}

\newcommand{\mathsc}[1]{{\normalfont\textsc{#1}}}
\newcommand{\PoA}{\mathsc{PoA}\xspace}
\newcommand{\IPoA}{\mathsc{I-PoA}\xspace}

\DeclareMathOperator{\E}{\mathbb{E}}

\newcommand{\R}{\mathbb{R}}
\renewcommand{\b}{\boldsymbol{b}}
\renewcommand{\v}{\boldsymbol{v}}

\newcommand{\pne}{\textsc{PNE}\xspace}
\newcommand{\mne}{\textsc{NE}\xspace}
\newcommand{\udb}{\textsc{UdB}\xspace}
\newcommand{\uni}{\textsc{Uni}\xspace}
\newcommand{\btv}{\textsc{BtV}\xspace}
\newcommand{\abe}{\textsc{AbE}\xspace}

\newcommand{\spa}{\textsc{SPA}\xspace}
\newcommand{\vcg}{\textsc{VCG}\xspace}
\newcommand{\fpa}{\textsc{FPA}\xspace}
\newcommand{\gsp}{\textsc{GSP}\xspace}

\newcommand{\lwel}{\textsc{Lwel}\xspace}

\newcommand{\rev}{\textsc{Rev}\xspace}

\newcommand{\eps}{\varepsilon}
\newcommand{\dd}{\mathrm{d}}

\title{{\bf Auto-bidding and Auctions in Online Advertising:\\
A Survey}\thanks{Authors' contacts: \texttt{\{gagana, ashwinkumarbv, sbalseiro, kshipra, dengyuan, zhef, gagangoel, cvliaw, haihao, mahdian, maojm, aranyak, mirrokni, renatoppl, perlroth,
gpil, jschnei, aschvartzman, balusivan, spendlove, yifengt, wadi, hanruiz, mingfei, wennanzhu, szuo\}@google.com}}}
\date{}

\author{Gagan Aggarwal}
\author{Ashwinkumar Badanidiyuru}
\author{Santiago R. Balseiro}
\author{Kshipra Bhawalkar}
\author{Yuan Deng}
\author{Zhe Feng}
\author{Gagan Goel}
\author{Christopher Liaw}
\author{Haihao Lu}
\author{Mohammad Mahdian}
\author{Jieming Mao}
\author{Aranyak Mehta}
\author{Vahab Mirrokni}
\author{Renato Paes Leme}
\author{Andres Perlroth}
\author{Georgios Piliouras}
\author{Jon Schneider}
\author{Ariel Schvartzman}
\author{Balasubramanian Sivan}
\author{Kelly Spendlove}
\author{Yifeng Teng}
\author{Di Wang}
\author{Hanrui Zhang}
\author{Mingfei Zhao}
\author{Wennan Zhu}
\author{Song Zuo}

\affil{Google}

\begin{document}

\maketitle

\begin{abstract}
  In this survey, we summarize recent developments in research fueled by the growing adoption of automated bidding strategies in online advertising. We  explore the challenges and opportunities that have arisen as markets embrace this autobidding and cover a range of topics in this area, including bidding algorithms, equilibrium analysis and efficiency of common auction formats, and optimal auction design.

\end{abstract}

\section{Introduction}
Autobidding systems are taking on an increasingly large role in the online advertising ecosystem, with strong adoption by advertisers. Traditional bidding interfaces require advertisers to submit fine-grained bids, e.g., one bid per collection of keywords. With autobidding, an advertiser submits a
high level goal and some high level constraints to the bidding platform. An {\em autobidding
  agent} for the advertiser then converts the goal and constraints into per-query bids
at auction time, based on predictions of performance
of each potential ad impression. Besides providing a much simplified interaction for the advertisers, autobidding also provides improved performance due to real-time optimization that takes predicted performance into account. Thus, it has already become a critical tool used by many advertisers.

There are several autobidding products in the advertising market. The oldest and most well-studied is that of budget optimization, in which the advertiser aims to maximize clicks from its ad campaign, and simply provides a daily budget and keywords to target. A more recent autobidding product is that of target-cost-per-acquisition (tCPA) in which the advertiser aims to maximize their post-click conversions (e.g., a sale or a sign-up), subject to an RoS (return-on-spend) constraint that the average cost per conversion is no more than a stated target. Target-return-on-ad-spend (tRoAS) generalizes tCPA to take into account the ultimate value of a conversion as well.

For each such product, the bidding agent automatically adjusts bids for its advertiser at auction time so as to maximize the performance for the campaign (under the given constraints), accounting for a dynamically changing environment such as query volume, query mix, or competition. We note that autobidding systems can be owned by the advertising platform as a service, or by third-parties. Given the importance of autobidding in the ad ecosystem, there has been considerable research in recent years to understand the fundamental properties, such as bidding algorithms, interaction with auction design, system equilibrium (i.e., the interaction across multiple autobidding agents for multiple advertisers), and mechanism design. This survey will attempt to cover a large portion of this growing and important literature.

\section{Preliminaries}\label{sec:prelim}

In this section, we define the problem faced by the autobidding agents and the auctioneer in a set of unified notations. Consider the environment with $n$ autobidding agents, indexed by $i$, and $m$ auctions, indexed by $j$. The valuation of agent $i$ winning auction $j$ is $v_{ij} \in \R_+$. In general, the auctions are heterogeneous such that $v_{ij}$ can vary with $j$. In the context of online advertising, the value $v_{ij}$ may include predictions from machine learning models such as predicted click-through-rate and/or predicted conversion-rate, which can be different from auction to auction depending on the auction features made available to the prediction models.

\paragraph*{Multi-slots}
In the multi-slot environment, each auction can have up to $\ell \geq 1$ slots, indexed by $k$. The $\ell$ slots under each auction have decreasing importance to the agents because the ones in lower positions are less likely to attract the attention of the end user. This is modeled by a decaying factor $\beta_{jk} \in [0, 1]$ decreasing in $k$, such that the agent $i$ winning the slot $k$ of the auction $j$ receives value $\beta_{jk} \cdot v_{ij}$. We assume that the decaying factor of the first slot is always normalized to $1$, i.e., $\beta_{j1} \vcentcolon = 1$. It is also without loss of generality to assume that each auction has the same number of slots, i.e., $\ell$, because for any auction with fewer slots, one can always by introducing virtual slots with decay factor $0$. To simplify the notations, in settings with only one slot, we will drop the decay factor $\beta_{jk}$.

\paragraph*{Bidding and Auction}
In each auction $j$, each agent submits a bid $b_{ij} \in \R_+$ and the auction takes the vector of bids $\b_j = (b_{1j}, \ldots, b_{nj})$ as the input and determines the allocations $x_{ij}(\b_j) \in [0, 1]$ and payments $p_{ij}(\b_j) \in \R$ for each agent $i$. In the settings with multiple slots, the allocation becomes $x_{ijk}(\b_j) \in [0, 1]$, indicating the potentially randomized allocation of the slot $k$ in auction $j$ to agent $i$.

\subsection{Bidding Agent's Problem}
\label{sec:bidding-agent-problem}
In the application of online advertising, the problem for the bidding agent is usually to maximize a given objective while subject to some constraints. There are two widely used objectives:
\begin{itemize}
  \item {\em Utility-maximizing objective}: $\sum_{j\in[m]} x_{ij} \cdot v_{ij} - p_{ij}$;
  \item {\em Value-maximizing objective}: $\sum_{j\in[m]} x_{ij} \cdot v_{ij}$.
\end{itemize}
Value maximization focuses on maximizing clicks or conversions, regardless of cost, appealing to advertisers who prioritize these metrics. It indirectly considers payments through a constraint on payments or the return on spend, which we discuss next. Utility maximization, common in economics, maximizes the difference between value and payments, requiring values to be expressed in monetary terms, which can be difficult for advertisers.
In certain settings, their hybrid version parameterized by $\lambda \in [0, 1]$ is also considered:
\begin{itemize}
  \item {\em Hybrid objective}: $\sum_{j\in[m]} x_{ij} \cdot v_{ij} - \lambda \cdot p_{ij}$.
\end{itemize}

The constraints in practice can be designed to implement many different features, such as guaranteeing the number of wins, or limiting the maximum bids, etc. Out of which, the most commonly studied constraints are the {\em budget} constraint and the {\em return-on-spend (RoS)} constraint.
\begin{itemize}
  \item {\em Budget constraint}: $\sum_{j\in[m]} p_{ij} \leq B_i$;
  \item {\em RoS constraint}: $\sum_{j\in[m]} x_{ij} \cdot v_{ij} \geq \tau_i \cdot \sum_{j\in[m]} p_{ij}$.
\end{itemize}
Budget constraints provide a natural way for advertisers to control their expenditures and are prevalent in advertising markets.
We note that the RoS constraint has many equivalent forms, such as target CPA (cost-per-action) constraint, ROI (return-on-investment) constraint, IR (individual rationality) constraint with scaled values, etc.
Throughout this survey, we will discuss all (equivalent) results in the language of the RoS constraints.

Taking advantage of the generality of the hybrid objective, we can formulate the bidding agent's problem as the following program:
\begin{align}
  \max &\qquad \sum_{j\in[m]} x_{ij} \cdot v_{ij} - \lambda \cdot p_{ij} \label{prog:bidding}\tag{\textsc{Bidding}}  \\
  \text{s.t.}& \qquad
    \sum_{j\in[m]} p_{ij} \leq B_i \label{prog:bidding-budget}\tag{\textsc{Budget}}  \\
    & \qquad
    \sum_{j\in[m]} x_{ij} \cdot v_{ij} \geq \tau_i \cdot \sum_{j\in[m]} p_{ij}  \label{prog:bidding-ros}\tag{\textsc{RoS}}
\end{align}
By picking different combinations of the parameters ($\lambda, B_i, \tau_i$), \eqref{prog:bidding} can capture most of the settings that we are interested in.
\begin{itemize}
  \item $\lambda = 0$: value-maximization;
  \item $\lambda = 1$: utility-maximization;
  \item $\tau_i = 0$: no RoS constraint;
  \item $B_i = +\infty$: no budget constraint.
\end{itemize}

\paragraph*{Research Questions} From the bidders' perspective, the most important questions are:
\begin{enumerate}
  \item {\em Optimal bidding} (Section~\ref{sec:bidding-truthful}): What is the optimal bidding strategy in a complete-information truthful auction?
  \item {\em Online bidding for truthful auctions} (Section~\ref{sec:bidding-online-truthful}): How should agents bid in online truthful auctions when competition and valuations are uncertain?
  \item {\em Online bidding for non-truthful auctions} (Section~\ref{sec:bidding-online-non-truthful}): How does non-truthfulness of the auction impacts agents' online learning strategies?
\end{enumerate}

\subsection{Auctioneer's Problem}

Similar to the welfare (or liquid welfare in the presence of budgets) and revenue maximization in the canonical auction design settings, (liquid) welfare and revenue are also the most commonly concerned properties in autobidding environments. One major difference, when the bidding agent's objective is value-maximization with RoS constraint, is that the revenue and the (liquid) welfare are the same as long as the budget constraint or the RoS constraint binds. More specifically, define
\begin{itemize}
  \item {\em Liquid welfare}: $\lwel = \sum_{i \in [n]} \min\{B_i, \sum_{j \in [m]} x_{ij} \cdot v_{ij} / \tau_i\}$;
  \item {\em Revenue}: $\rev = \sum_{i \in [n]} \sum_{j \in [m]} p_{ij}$.
\end{itemize}
Liquid welfare is a measure of efficiency introduced by \citet{dobzinski2014efficiency}, which quantifies the highest possible revenue that can be attained by a seller with full information on the bidders' information. In the literature, liquid welfare is used to measure efficiency instead of the usual social welfare because the later cannot be well-approximated when bidders are constrained.

Observe that for each $i$,
\begin{gather*}
  \sum_{j \in [m]} p_{ij}
  \leq \min\{B_i, \sum_{j \in [m]} x_{ij} \cdot v_{ij} / \tau_i\},
\end{gather*}
therefore $\rev \leq \lwel$. The equality is reached when for all agent $i$, either \eqref{prog:bidding-budget} or \eqref{prog:bidding-ros} binds.

For value-maximization agents, i.e., $\lambda = 0$, under their optimal strategy (assuming others' fixed), it is often the case that either \eqref{prog:bidding-budget} or \eqref{prog:bidding-ros} binds, unless there is no chance for them to obtain higher values by increasing their spend. For this reason, (liquid) welfare receives significantly more attention in the literature, especially with value-maximization agents.

\paragraph*{Research Questions}
From the auctioneer's perspective, roughly three major categories of problems are concerned:
\begin{enumerate}
  \item {\em Equilibrium} (Section~\ref{subsec:equilibrium}): Does an equilibrium exist and, if so, it is unique and can it be efficiently computed? Do bidders converge to an equilibrium under different dynamics?
  \item {\em Price of anarchy} (PoA, see Section~\ref{subsec:poa} for formal definition): How are efficient are  equilibria in autobidding auctions? How auction design impacts the price of anarchy?
  \item {\em Optimal auction design} (Section~\ref{sec:auction}): How can we design auctions that improve the revenue or efficiency of the market?
\end{enumerate}

\subsection{Bayesian Auction Model}\label{subsec:bayesian-model}

An alternative model widely adopted in the literature is a Bayesian model with a single auction in which bidder's values are drawn from a distribution $F \in \Delta (\mathbb R_+^n)$. (For simplicity, we discuss the case of a single-slot auction.) The enumerative model defined above can be represented as a Bayesian model in which the random valuations  $\tilde \v = (\tilde v_1, \ldots, \tilde v_n)$ are drawn from a distribution $F$ that takes value $(v_{1j}, \ldots, v_{nj})$ with probability $1/m$ for each $j \in [m]$. Bayesian models are prominent in optimal auction design as they are useful to encode different informational assumptions and to represent settings with a continuum of valuations, which sometimes lead to more analytically tractable models.

Table~\ref{tab:model-comp} summarizes the key elements under the two different models. In the Bayesian model the budget and RoS constraints are naturally written in expectation over the realization of valuations and any randomness of the mechanism. This model can be interpreted as single-period problem with expected value constraints, instead of the more complex real-world scenario of multiple auctions with average constraints over time. This approach is more manageable and oftentimes allows for explicit solutions.

\begin{table}[]
  \centering
  \begin{tabular}{c|l|l}
    & $m$-auction model & Bayesian model  \\
    \hline
    Hybrid objective & $\sum_{j\in[m]} x_{ij} \cdot v_{ij} - \lambda \cdot p_{ij}$ & $\E[x_i \cdot v_i - \lambda \cdot p_i]$ \\
    \hline
    Budget constraint & $\sum_{j\in[m]} p_{ij} \leq B_i$ & $\E[p_i] \leq B_i / m$ \\
    \hline
    RoS constraint & $\sum_{j\in[m]} x_{ij} \cdot v_{ij} \geq \tau_i \cdot \sum_{j\in[m]} p_{ij}$ & $\E[x_i \cdot v_i] \geq \tau_i \cdot \E[p_i]$ \\
    \hline
    Liquid welfare & $\sum_{i \in [n]} \min\{B_i, \sum_{j \in [m]} x_{ij} \cdot v_{ij} / \tau_i\}$ & $\sum_{i \in [n]} \min\{B_i / m, \E[x_i \cdot v_i] / \tau_i\}$ \\
    \hline
    Revenue & $\sum_{i \in [n]} \sum_{j \in [m]} p_{ij}$ & $\sum_{i \in [n]}\E[p_i]$
  \end{tabular}
  \caption{Comparison between two models.}
  \label{tab:model-comp}
\end{table}

\section{Bidding Algorithms}\label{sec:bidding}

\subsection{Optimal bidding for truthful auctions}\label{sec:bidding-truthful}

In this section we present an LP based formulation of the autobidding problem from the view of an agent for one advertiser, slightly extending~\citet{aggarwal2019autobidding} to account for hybrid objectives.

Throughout this section, we will omit the subscript of $i$ as we will take the perspective of a single bidding agent while assuming all other agents are fixed.
Let $p_j$ be the price of an ad for this advertiser for query $j$. Clearly, $p_j$ depends on the bids of the other advertisers (who may also be using bidding agents), as well as on the pricing rule of the underlying auction.
Suppose, for argument, we know the query sequence and the values of $p_j$ in advance. Then we have the \emph{selection problem} from Section~\ref{sec:bidding-agent-problem}: which queries would the advertiser like to buy so as to maximize their objective while staying within their constraints.

We rewrite the LP together with its Dual LP below, for a more generalized family of bidding problems with multiple constraints capturing different budget and RoS constraints $c\in \mathcal{C}$, parameterized by $B^c$, and $w_j^c$. The RoS constraint (like tCPA) is captured by setting the corresponding $B^c = 0$ and $w_j^c = \tau\cdot v_j$, and a budget constraint is captured by setting the corresponding $B^c$ as the budget and $w_j^c = 0$. Finally, we remark that the extremal solutions of the LP are mostly integral if the query stream is large. \\
\begin{minipage}{0.47\textwidth}
\small
\begin{align}
    \mbox{Maximize  } \textstyle \sum_j v_j x_j &- \lambda p_j x_j \mbox{~~s.t.}\label{eq:tcpa-obj-gen}\\
    \forall c\in \mathcal{C}: \textstyle \sum_j p_j x_j &\leq B^c +  \textstyle \sum_j w_j^c x_j \notag\\
    x_j &\leq 1 \label{eq:tcpa-gen-x-constraint}\\
    x_j &\geq 0 \notag
\end{align}
\end{minipage}
\hfill\vline\hfill
\begin{minipage}{0.52\textwidth}
\small
\begin{align}
    &\mbox{Minimize  } \textstyle \sum_j \delta_j + \textstyle \sum_c \alpha_c B^c ~~\mbox{s.t.}\label{eq:tcpa-obj-dual-gen}\\
    &\forall~ j: \delta_j \geq \textstyle \sum_c \alpha_c (w_j^c - p_j) + v_j - \lambda p_j \label{eq:tcpa-constraint-dual-gen} \\
    &\forall~j: \delta_j \geq 0 \notag \\
    &\forall~ c\in \mathcal{C}: \alpha_c \geq 0 \notag
\end{align}
\end{minipage}

\vspace{1em}
In the dual problem, we denote by $\alpha_c$ the dual variable of the constraint $c \in \mathcal C$ and $\delta_j$ the dual variable of the constraint \eqref{eq:tcpa-gen-x-constraint}. We now leverage the LP formulation to come up with a bidding formula which can achieve the same optimal choice of queries as in the selection problem.
The dual constraint (\ref{eq:tcpa-constraint-dual-gen}) can be re-written as:
\begin{equation}
\label{eq:tcpa-dual-gen-rewrite}
\forall ~j: \frac{\delta_j}{\lambda + \sum_c \alpha_c} \geq \left(\frac{v_j + \sum_c \alpha_c w_j^c}{\lambda + \sum_c \alpha_c} - p_j\right)
\end{equation}

This directly gives us a bidding formula: Set the bid for query $j$,
\begin{equation}
\label{eq:optimal-bid}
b_j := \frac{v_j + \sum_c \alpha_c w_j^c}{\lambda + \sum_c \alpha_c}
\end{equation}

\begin{theorem}\label{thm:uniform-bidding}
Assuming that we have access to optimal values of the dual variables $\alpha_c$, the bidding formula (\ref{eq:optimal-bid}) results in an auction outcome identical to an optimal primal solution $x_j$, if the underlying auction is truthful.
\end{theorem}

For a value maximizer ($\lambda = 0)$ and simple RoS constraint with an additional a budget constraint (with corresponding optimal dual variables $\alpha_T$ and $\alpha_B$, the bidding formula becomes:
\begin{equation}\label{eq:uniform-bidding}
b_j = \left(  \frac{1 + \tau\cdot \alpha_T}{\alpha_T + \alpha_B}\right) v_j
\end{equation}
For only an RoS constraint, the formula becomes $b_j = \left(\frac{1}{\alpha_T} + \tau\right)v_j$, essentially bidding proportionally to the value for an appropriate constant of proportionality. Finally, for a utility maximizer ($\lambda=1)$ with a budget constraint, we obtain the bidding formula
\begin{equation}\label{eq:uniform-bidding-utility}
b_j = v_j / (1+\alpha_B)
\end{equation}
that was introduced by \cite{balseiro2015repeated}.

The bid formula depends on the knowledge of the optimal duals $\alpha_c$; in practice these can be estimated via ML techniques from past data, and updated via control loops. We discuss the design of online algorithms in the next subsection.

\subsection{Online learning for bidding in truthful auctions}\label{sec:bidding-online-truthful}

There is a recent line of work studying the design of online learning algorithms under uncertainty. Most work in the literature consider a finite horizon model in which the bidder participates in $m$ sequential auctions and constraints are enforced over time across auctions. For simplicity, we consider a single-slot auction. In this online model, when the $j$-th auction is announced, features are shared with the bidder and they estimate a value $v_j$ for winning the auction. The value $v_j$ is exogenously given and usually estimated using offline machine learning models. Valuation models tend to be more stable over time and can be trained across many advertisers and long periods of time~\citep{mcmahan2013ad,he2014practical,zhou2018deep,juan2016field,lu2017practical}. The payment $p_j$ is learned after the auction is cleared. While the value $v_j$ is learned before bidding in the $j$-th auction, future values are not known in advance.

The online learning problem has been studied under different input models, stochastic and adversarial, and for truthful and non-truthful auctions. The case of non-truthful auctions is notoriously harder because the payment depends on the bid and the uniform bidding formula in \eqref{eq:uniform-bidding} is not optimal---instead, the optimal bidding formula is a non-linear function of the value. We consider the case of non-truthful auctions in the next subsection.

There is a line of work studying the bidding dynamics and resulting market efficiency when all agents simultaneously adopt learning algorithms, which we summarize in Section~\ref{sec:bidding-dynamics}.

\paragraph*{Stochastic input} For truthful auctions, it is commonly assumed that the pairs $(v_j,p_j) \sim \mathcal P$ are  independently and identically distributed (i.i.d.) from a joint distribution $\mathcal P$ that is unknown to the bidder. In other words, values and payments can be arbitrarily correlated for a given auction but independent across auctions. In this case, dual-based algorithms that update the dual variables $\alpha_B$ and $\alpha_T$ for the budget and RoS constraints, respectively, using first-order algorithms from online optimization has been shown to attain low regret relative to the offline optimization problem \eqref{prog:bidding}.

\cite{balseiro2019learning} consider the problem of a utility maximizer ($\lambda = 1$) with only a budget constraint ($\tau = 0$) and proposed \emph{dual gradient descent}, a simple algorithm that adjusts the dual variable iteratively using online gradient descent. Denoting by $\alpha_{B,j}$ the value of the dual variable at the beginning of auction $j$, following \eqref{eq:uniform-bidding-utility}, the bid is set to $b_j = v_j/(1+\alpha_{B,j})$. Initially, the dual variable is set to $\alpha_{B,0} = 0$ and then it is updated as follows
\begin{align}
    \alpha_{B,j+1} = \max\left( \alpha_{B,j} + \eta\cdot(B / m - p_j), 0\right)\label{eq:dual-update}
\end{align}
where $\eta$ is a step-size that is usually chosen to be of order $\eta \approx m^{-1/2}$. Budget constraints are usually enforced strictly and the algorithm stops bidding whenever the budget is depleted. The term $B / m - p_j$ can be shown to be a subgradient of the $j$-th term of the dual objective \eqref{eq:tcpa-obj-dual-gen} so the algorithm can be interpreted as performing gradient descent in the dual problem.
The algorithm has an appealing self-correcting feature: it adjusts the dual variable to guarantee that the spend per auction is close to the average spend $B / m$.
\citet{balseiro2019learning} show this algorithm obtains the following regret guarantee
\[
    \sup_{\mathcal P} \E_{(v_j,p_j)\sim \mathcal P} \left[ \mathrm{OPT} - \mathrm{ALG} \right] = O(\sqrt{m})\,,
\]
where $\mathrm{OPT}$ denotes the optimal objective value of the offline bidding problem \eqref{prog:bidding}. The result of \citet{balseiro2019learning} is proven under restrictive assumptions on the distributions $\mathcal P$, which were later relaxed by \citet{balseiro2023best}.

\cite{feng2023online} provide a dual-based algorithm for a value maximizer with a budget and RoS constraint.
They prove their algorithm attains regret $O(\sqrt{m})$ and incurs a violation of at most $O(\sqrt{m} \log m)$ of the RoS constraint. They also provide a more refined algorithm that satisfies both constraints strictly.

The near-optimal bidding algorithm of \citet{feng2023online} requires coordination between budget and RoS pacing systems to determine the bid.
\citet{balseiro2023joint} explore algorithms with different degrees of coordination between pacing systems. In particular, they show that a fully-decoupled sequential algorithm could lead to poor performance and constraints violations, while a minimally-coupled algorithm that runs services independently can achieve similar performance to the optimal, fully-coupled algorithm.

\paragraph*{Adversarial input} When values and payments are adversarially chosen, it is generally not possible to obtain fixed competitive ratios and, instead, one should settle for data-dependent competitive ratios. For value and utility maximizers with a budget constraint, \citet{zhou2008budget} provide a dual-based algorithm that bids according to \eqref{eq:uniform-bidding-utility} and updates the dual variable based on the budget spent. Their algorithm attains a near-optimal competitive ratio of $1 + \log(U/L)$ where $L$ and $U$ are lower and upper bounds, respectively, on the value/utility to payment ratios.
Their proof is established under the so-called ``small bids assumption'' that requires payments to be small relative to the budget.
Later, \cite{balseiro2019learning} show that dual gradient descent obtains a competitive ratio of $\sup_j v_j / (B/m)$ for utility maximizers with a budget constraint. Their competitive ratio is also shown to be tight, which makes the algorithms of \citet{zhou2008budget} and \citet{balseiro2019learning} not directly comparable.

\subsection{Online learning for bidding in non-truthful auctions}\label{sec:bidding-online-non-truthful}

When the auction is non-truthful, Theorem~\ref{thm:uniform-bidding} does not hold and the optimal bid can be a complex function of values. A na\"ive approach is to reduce the problem to a contextual bandit with knapsacks in which the context is the value and each arm is a bid~\citep{badanidiyuru2014resourceful}. This approach requires discretization and results in a suboptimal regret bound of $O(m^{2/3})$ as it fails to exploit the structure of the problem. The most prominent non-truthful auction studied in the literature is the first-price auction in which the highest bidder wins and pays their value. First-price auctions have been recently adopted by many advertising platforms: Google switched to first-price auctions for its ad exchange in 2019 and Twitter switched in 2020 for mobile apps.\footnote{See \url{https://www.blog.google/products/admanager/rolling-out-first-price-auctions-google-ad-manager-partners} and \url{https://www.mopub.com/en/blog/first-price-auction}.}

\paragraph*{First-price auctions with budgets}
\cite{wang2023learning} focus on stochastic inputs where the values and the maximum competing bids are drawn from two independent distributions. For the full feedback model where the maximum competing bid is revealed after every auction, they provide a primal-dual algorithm that attains $\tilde{O}(\sqrt{m})$ regret. Using the graph-feedback and partial order properties in first-price auctions identified in \cite{han2024optimal}, they also provide an algorithm with $\tilde{O}(\sqrt{m})$ regret in the one-sided feedback model where the bidders observes the maximum competing bid only if they lose the auction. \citet{ai2022no} study a model that additionally involves a discount factor in the objective function. They show $\tilde{O}(\sqrt{m})$ regret with full feedback and $\tilde O(m^{7/12})$ with one-sided feedback. \cite{castiglioni2022online} provide a general algorithm   for the online knapsack problem with multiple resource constraints that can lead to low regret guarantees for stochastic inputs. Their algorithm is based on the Lagrangian framework of \cite{immorlica2022adversarial}. As an application of their framework, they discuss the problem of bidding in first-price auctions with budget constraints.

\paragraph*{First-price auctions with budget and RoS constraints}
\citet{castiglioni2024online} solve the general online learning problem under budget and RoS constraints. They endow standard primal-dual templates with weakly adaptive regret minimizers.
Their framework applies to repeated first-price auctions where the set of possible valuations and bids are finite.
\citet{aggarwal2024no} solve the problem of continuous valuations by designing an algorithm under full-information feedback, with $\tilde O(\sqrt{m})$ regret against the best possible Lipschitz function that maps values to bids.
Their result builds on the primal-dual framework in \citet{castiglioni2024online} and is obtained by designing a dedicated tree-structured primal regret minimizer that achieves low interval regret.
They also provide a lower bound of $\Omega(m^{2/3})$ regret with bandit feedback.
\citet{liang2023online} design learning algorithms for an advertiser who repeatedly interacts with a platform when the selling mechanism/autobidding algorithm is a black box. They present a primal-dual algorithm for bandit feedback that attains good performance under different input models.

\section{Equilibria and Price of Anarchy}\label{sec:poa}

Within this section, most of the results only apply to the value-maximizing objective (i.e., $\lambda = 0$), and therefore, without explicit note, we assume $\lambda = 0$.

\subsection{Solution Concepts}
\label{subsec:solution_concepts}
After defining the action space and the objective function for the agents, one natural question is to understand the game in terms of the properties of equilibria as well as other extended solution concepts. We summarize some solution concepts studied in the literature. We note that when either \eqref{prog:bidding-budget} or \eqref{prog:bidding-ros} is violated, the objective value for the agent is defined as $-\infty$. So in all the solution concepts below, \eqref{prog:bidding-budget} and \eqref{prog:bidding-ros} are forced to be satisfied by all agents.

\begin{itemize}
  \item {\em Nash equilibrium} (\mne): No agents can improve its objective value by changing to a different action $\b'_i$ within the set of (randomized) bids, i.e., $\b'_i \in \Delta(\R_+^m)$.
  \item {\em Pure Nash equilibrium} (\pne): No agent can improve its objective value by changing to a different action $\b'_i$ within the set of deterministic bids, i.e., $\b'_i \in \R_+^m$.
  \item {\em Undominated bids} (\udb) \cite{balseiro2021robust}: No agent chooses a {\em dominated action}, i.e., $\b_i \in \udb_i$. An action $\b_i$ dominates another action $\b'_i$, if (i) For all possible bid profiles from others $\b_{-i} = (\b_1, \ldots, \b_{i-1}, \b_{i+1}, \ldots, \b_{n})$, agent $i$'s objective value from $(\b_i, \b_{-i})$ is weakly higher than that from $(\b'_i, \b_{-i})$; (ii) And there exists one bid profile from others $\b''_{-i}$, such that agent $i$'s objective value from $(\b_i, \b''_{-i})$ is strictly higher than that from $(\b'_i, \b''_{-i})$.
  $\udb_i$ is the set of actions of agent $i$ that is not dominated by any other action.
  \item {\em \pne within \udb} ($\pne+\udb$): All agents choose {\em undominated} actions and no agent can improve its objective value by changing to a different {\em undominated} action, i.e., $\b'_i \in \R_+^m \cap \udb_i$.
  \item {\em \pne within Uniform Bidding} ($\pne+\uni$): All agents choose {\em uniform-bidding} actions and no agent can improve its objective value by changing to a different {\em uniform-bidding} action $\b'_i \in \uni_i = \{\alpha \cdot \v_i : \alpha \in \R_+\}$.
  \item {\em Better-than-bidding-Values} (\btv) \citep{deng2022efficiency}: No agent can improve its objective value by changing to the action of directly bidding its values, $\b'_i = \v_i$.
  \item {\em Autobidding Equilibrium} (\abe) \citep{li2024vulnerabilities} (Second-price auction only): An \abe consists of the uniform-bidding actions for each agent and the allocation of the auctions such that: (i) Only agents with the highest bids can have non-zero allocation for the corresponding auctions, and all auctions are fully allocated; (ii) With the second price rule, the RoS constraints are respected and moreover, must bind unless the corresponding uniform-bidding factor reaches the upper limit.
\end{itemize}

We note that the relationship between these solution concepts could be complicated, and sometimes depends on the format of the auction. For example, \pne+\udb by definition is a subset of \pne, while \pne+\uni is not. In contrast, \btv is a necessary condition for a Nash equilibrium.

This section describes price of anarchy (PoA) results for a number of different solution concepts as described above. However, we note that our definition of PoA may not be ``standard'' since we may impose additional constraints that limit what equilibria are possible. Nonetheless, these results are useful from a practical point of view since the additional assumptions (such as \udb and \uni) are mild in practice.

\subsection{Equilibrium existence and complexity}\label{subsec:equilibrium}

\citet{aggarwal2019autobidding} shows the existence of PNE among autobidding agents in general multi-slot truthful auctions with two mild technical assumptions.
One of them is, essentially, that there is no point mass in the value distributions. This assumption
can be avoided by incorporating appropriate tie-breaking into the solution concept; this is an important advancement as a large body of the autobidding literature studies discrete instances, i.e., the values $v_{ij}$ are deterministic and both $n$ and $m$ are finite. \citet{conitzer2022budgetpacing} introduce this in the context of budget-constrained bidders and define the notion of a second-price pacing equilibrium (SPPE) -- an SPPE is characterized by a vector of pacing multipliers as well as a fractional allocation of tied impressions that satisfies all the budget constraints. They prove the existence of SPPE for every pacing game, including those that are discrete and discontinuous, by constructing a sequence of smoothed games that converge to the (non-smoothed) pacing game. They show that a PNE exists for each of the smoothed games and the sequence of PNEs converges to an SPPE of the pacing game. \citet{li2024vulnerabilities} extend this definition to RoS-constrained bidders, defining the term \abe, and also give a similar construction to prove existence of \abe for RoS-constrained bidders.

Another alternative to proving existence of an equilibrium is assuming a continuum of values so that ties are a zero-probability event. This approach has been successfully applied to utility-maximizers with budget constraints bidding in truthful auctions when values are independent or correlated but with positive densities and non-truthful standard auctions such as first-price auctions (see, e.g., \citealt{balseiro2015repeated, balseiro2021budget,balseiro2023contextual}).

\citet{chen2023complexity} study the complexity of computing an equilibrium of the pacing game for budget-constrained bidders, and show that it is PPAD-hard to compute. \citet{aggarwal2023multi} extend their result to show that it is PPAD-hard to compute an \abe for RoS-constrained bidders. \citet{li2024vulnerabilities} show that finding an approximate-\abe is also PPAD-hard.

\subsection{Price of anarchy under different per-item auctions}\label{subsec:poa}

Since \citet{aggarwal2019autobidding}, there are several lines of work that focus on the price of anarchy of canonical auction formats as well as their variants. Many of them consider the case with $\lambda = 0$ and are subject to the \eqref{prog:bidding-ros} constraint. Some of them consider the \eqref{prog:bidding-budget} constraint in addition.

To begin with, we first formally define the notion of {\em price of anarchy} (a.k.a. PoA) with respect to a solution concept.
Let $\cI$ denote the set of all autobidding environment instances, and $I \in \cI$ denote one such instance. Let $\cA$ denote an auction format, such as Vickrey-Clarke-Groves (VCG) auction, first price auction (FPA), etc. Let $E$ be the solution concept and $E(\cA, I)$ be the corresponding set of bid profiles under auction format $\cA$ and instance $I$ that satisfy the solution concept $E$.

\begin{definition}[Price of Anarchy]
\label{def:poa}
The price of anarchy of an auction $\cA$ with respect to solution concept $E$ is given by
\[
    \PoA_E(\cA) = \sup_{I \in \cI,~\b \in E(\cA, I)} \frac{\max_{x^*}\lwel(x^*)}{\lwel(x^\cA(\b))},
\]
\end{definition}
where $x^\cA$ is the allocation function of the auction format $\cA$, and the $\max$ is taken over all allocations $x^*$ that are feasible with respect to the constraints. Note that here both $x^*$ and $\lwel$ depend on the instance $I$.

At a high-level, the price of anarchy tells us how much worse the worst-case equilibrium is from an optimal centralized allocation.
The price of anarchy is always at least $1$ and the closer it is to $1$ the better.

In the above definition, ``equilibrium'' broadly refers to a solution concept detailed in Section~\ref{subsec:solution_concepts}.
The specific concept used in each context will be clarified in the relevant section.

\subsubsection{Basic auction formats: SPA, VCG, FPA, and GSP}

The \PoA result by \citet{aggarwal2019autobidding} implies that in a second price auction, $\PoA_{\pne+\uni}(\spa) \leq 2$. Furthermore, they show that this \PoA bound is tight by showing that for $\eps > 0$, there is an instance such that $\PoA_{\pne+\udb+\uni}(\spa) \geq 2 - \eps$. The original theorem is in fact for a much more general setting, where each bidding agent is subject to multiple general constraints that cover both \eqref{prog:bidding-budget} and \eqref{prog:bidding-ros} as special cases. The solution concept is then \pne plus the optimal bidding strategy based on the dual variables of the constraints, which degenerates to a uniform bidding strategy when one only has \eqref{prog:bidding-budget} and \eqref{prog:bidding-ros} as the constraints. A generalized version of liquid welfare is also used for defining the notion of \PoA.

\medskip

\citet{deng2021towards} generalize the bound of $2$ to multi-slot settings, and hence proves that $\PoA_{\pne+\uni}(\vcg) \leq 2$. This bound is tight as \spa can be considered as \vcg on special single-slot instances. Hence $\PoA_{\pne+\uni}(\vcg) \geq \PoA_{\pne+\uni}(\spa) = 2$. They also extend the results to \vcg with certain additive boosts, which we will cover in Section~\ref{subsubsec:reserve-boost}.

\medskip
Beyond truthful auctions, when there is only \eqref{prog:bidding-ros} but no \eqref{prog:bidding-budget}, \citet{liaw2023efficiency} proves that $\PoA_{\pne}(\fpa) = 2$, and
\citet{deng2022efficiency} generalizes the result to randomized strategies, i.e., $2 \leq \PoA_{\mne}(\fpa) \leq \PoA_{\btv}(\fpa) \leq 2$. When there are both agents with $\lambda = 0$ and $\lambda = 1$ in the environment, \citet{deng2022efficiency} further shows that $\PoA_{\mne}(\fpa) = 1 + \max_{t \in [0, 1]} \frac{1 - t}{1 + t \ln t} \approx 2.188$. $\PoA_{\mne}(\fpa)$ can be improved with proper reserve prices, which we will cover in Section~\ref{subsubsec:reserve-boost}.

\medskip
As a non-truthful generalization of \spa in multi-slot settings, \gsp is another  auction format commonly studied in the literature. \citet{deng2023efficiency} proves an upper bound for $\PoA_{\btv}(\gsp)$ depending on the decaying factors $\beta_{jk}$, which is unbounded in general after taking $\sup$ over all possible instances, i.e., $\PoA_{\btv}(\gsp) = \infty$.  \citet{deng2023autobidding} further show a fined-grained PoA with respect to the discount factors, i.e., the ratios of click probabilities between lower slots and the highest slot in each auction.

\subsubsection{Basic auction formats with reserves and additive boosts}
\label{subsubsec:reserve-boost}

A line of research studies how PoA can be improved when the auctioneer has additional information about agent values and uses it as simple adjustments on basic auction formats.

\citet{deng2021towards} shows that in position auctions with value maximizing agents with \eqref{prog:bidding-ros} constraints (and potentially also \eqref{prog:bidding-budget} constraints), for any constant $c > 0$, if the auctioneer sets an additive boost for each agent as $c$ times the agent's value in VCG, $\PoA_{\pne+\uni}$ is at most $(c+2) / (c+1)$. As $(c+2) / (c+1) < 2$ for any $c>0$, this is a strict PoA improvement compared to VCG, and this PoA approaches 1 when $c$ goes to infinity.

\citet{balseiro2021robust} improve upon \citet{deng2021towards} in mainly three directions: (1)
allowing auctioneer's additional information (a.k.a. ML advice) about agent values to be approximate, (2) studying reserves in addition to additive boosts, and (3) considering the mixed environment with general agent for $\lambda \in [0, 1]$. In particular,  they show that if the auctioneer has a signal $\in [\gamma \cdot \text{value}, \text{value})$ for each agent, VCG with reserves or additive boosts gets $\PoA_{\udb}$ at most $2-\gamma$, and VCG with both reserves and additive boosts gets PoA at most $2 / (1+ \gamma)$. They also extend these results from VCG to GSP with slightly weaker guarantees.
With such ML advice, \citet{deng2022efficiency} also show this result can be extended to FPA with reserves setting to obtain $\PoA_\mne  = \min_{t \in [0, 1]}$ $\frac{2 - t - \gamma + (1 - \gamma) t \ln t}{1 + t \ln t - \gamma t (1 + \ln t)}$.

In the presence of user costs, \citet{deng2023autobidding} observe that the PoA of \vcg can be arbitrarily bad (i.e., $\PoA_{\pne+\udb}(\vcg) = \infty$). They show that constant PoA can be restored by introducing either auction-dependent reserve prices or agent-dependent reserve prices.

\subsubsection{Basic auction variants with randomization}

The first paper to study randomized auctions in the autobidding setting is \citet{mehta2022auction}.
The mechanism (RAND) considered in this paper is defined by two parameters: a gap parameter $\alpha$ and a swap probability $p \leq 1/2$.
If the gap between the highest bidder and the second highest bidder is at least $\alpha$ then the highest bidder wins. Otherwise, the highest bidder wins with probability $p$ and with the remaining probability, the second highest bidder wins.
The payment for the bidders is computed using Myerson's payment rule.\footnote{Fix a bidder $i$ and let $x_i(b_i, b_{-i})$ be the allocation to bidder $i$ when their bid is $b_i$ and all other bids are $b_{-i}$. Myerson's payment rule says that the payment to bidder $i$ should be given by $b_i x_i(b_i, b_{-i}) - \int_{0}^{b_i} x(t, b_{-i}) \, \dd t$.}
They show that in the setting with two bidders, there is a choice of $\alpha \geq 1$ and $p$ that gives a PoA of around $1.9$. Note that even when there are only $2$ bidders, the example in \citet{aggarwal2019autobidding} shows that the PoA of SPA is $2$.

A followup work by \citet{liaw2023efficiency} considers mechanisms that are both randomized and non-truthful (i.e., the payment does not necessarily follow Myerson's payment rule).
More specifically, they consider a mechanism called randomized first-price auction (rFPA), which has an allocation function which is a generalization of RAND, but charges each winning bidder its bid. They show that, with an appropriate choice of their parameter $\alpha$, this further improves the PoA to $1.8$.
We note that a key difference between \citet{liaw2023efficiency} and \citet{mehta2022auction,aggarwal2019autobidding} is that since the auction is no longer truthful, uniform bidding is not a best response and one has to analyze all possible bidding strategies.

We note that it is an open problem to design a randomized mechanism that has a PoA of \emph{strictly} less than $2$ for any fixed number $n$ of bidders. A more difficult open problem is to exactly compute the PoA as a function of $n$. We remind the reader that these open problems are in the setting where the auction only receives bids and does not have any prior information on the values.

\subsubsection{In the presence of budget constraint}

\citet{liaw2024efficiency} studies the autobidding settings with both \eqref{prog:bidding-budget} and \eqref{prog:bidding-ros}. They first show that the gap between the optimal deterministic allocation and the optimal randomized allocation is $n$. Next, they define integral-PoA ($\IPoA$), which is the same as Definition~\ref{def:poa}, with an extra constraint that $x_{i,j}^* \in \{0, 1\}$. They show that the $\PoA$ of FPA is $n$, but it decreases to 2 under the mild assumption that for any bidder, their value for any query is at most their total budget. Interestingly, the $\IPoA$ of FPA is 2 when there is only a single \eqref{prog:bidding-ros} constraint \citep{liaw2023efficiency} and when there are both \eqref{prog:bidding-budget} and \eqref{prog:bidding-ros} constraints. This means that the $\IPoA$ does not get worse when the \eqref{prog:bidding-budget} budget constraint is added on top of the \eqref{prog:bidding-ros} constraint.

Uniform bidding is shown to be near optimal for bidders in truthful auctions \citet{aggarwal2019autobidding}, and achieves an optimal $\PoA$ of 1 for FPA with \eqref{prog:bidding-ros} constraints \citet{deng2021towards}. \citet{liaw2024efficiency} shows that the $\IPoA$ of FPA with uniform bidding is $n$, which is worse than the $\IPoA=2$ for non-uniform bidding. The reason is that the bidders could be in a situation that they either win no query or would violate their budget by uniformly increasing bids for every query. However, uniform bidding improves the $\PoA$ for rFPA, because the bidders could increase bids smoothly to get more fraction value to avoid the bad cases in deterministic auctions. Finally, the authors propose a “quasi-proportional” FPA mechanism that achieves a $\PoA$ of 2 with both \eqref{prog:bidding-budget} and \eqref{prog:bidding-ros} constraints.

\subsection{Bidding Dynamics}\label{sec:bidding-dynamics}

Even though equilibria are shown to exist under some conditions, it remains unclear whether the bidding agents will eventually converge to an equilibrium by following their bidding algorithms.

\citet{borgs2007dynamics} study the dynamics of budget constrained agents bidding in second-price auctions and first-price auctions under uniform bidding. They prove convergence for first-price auction when bids are randomly perturbed and numerically explore the dynamics under second-price auctions. \citet{balseiro2019learning} study utility-maximizing agents with budget constraints bidding in second-price auctions and show that dynamics converge to a unique equilibrium when the expected expenditure of bidders satisfy a strong monotonicity condition.

The work of \citet{paesleme2024complex} shows that even with simple bidding algorithms, complex behavior can emerge in autobidding systems.  For example, in one case of two bidders, there can be bi-stability (i.e., the existence of two stable equilibria), and the equilibrium to which the bidders converge depends upon the initial configuration of the multipliers.  In the case of three bidders, they observe that there can be a stable periodic orbit, which implies that for some initial conditions the bidding system will never converge, even if an equilibrium does exist.
Furthermore, they show that autobidding systems can simulate both linear dynamical systems as well as logical Boolean gates.

\citet{liu2023auto} study the optimal bidding strategy as the response to fixed strategies from competing agents in second price auctions. All agents have utility-maximization objectives under both budget and RoS constraints. When all agents adopt the proposed response strategy, they provide a sufficient condition such that the bidding dynamics converge to an equilibrium.

The line of work pioneered by \citet{gaitonde2022budget} studies the market efficiency when bidding agents simultaneously adopt learning algorithms. They show the liquid welfare obtained when all autobidders adopt the gradient-based algorithm in \eqref{eq:dual-update} is at least half of the optimal liquid welfare. Remarkably, their result does not require existence of an equilibrium, nor convergence of the dynamics. In their paper, they study utility-maximizing agents with stochastic values under budget constraints bidding in  second-price auctions or other auction formats such as first-price auction when bidders are restricted to uniform bidding. \citet{lucier2024autobidders} show a similar PoA guarantee of two for value-maximizers with budget and RoS constraints.

\citet{fikioris2023liquid} study the efficiency of value-maximizing bidders with budget constraints when values are adversarial. If every agent adopts an algorithm that guarantees a competitive ratio of $\gamma \ge 1$ compared to the best uniform bidding streategy, then the PoA of liquid welfare is $\gamma + 1/2 + O(\gamma^{-1})$. A remarkable feature of their result is that agents can adopt different algorithms. When $\gamma = 1$, their analysis yields a PoA of 2.41.

\section{Auction design}\label{sec:auction}

Given the behavior of bidding agents defined by \eqref{prog:bidding}, a natural question is what are the efficient (optimal) auctions. From Section~\ref{sec:poa}, we know that most of the commonly studied auction mechanisms are approximately efficient with constants at least $1.8$.
In this section, we introduce recent works on optimal auction design where the agents follow the optimization problem \eqref{prog:bidding}.

\subsection{Bayesian auction design}
\label{subsec:bayesian}

In this subsection, we focus on the single Bayesian auction model introduced in Section~\ref{subsec:bayesian-model}, which is general enough to capture the discrete $m$-auction model.

In general, each agent $i$ has three types of private information: (i) value $v_i$, (ii) budget $B_i$, and (iii) RoS target $\tau_i$. Table~\ref{tab:auctions} classifies the recent works based on whether each of these information is private or public as well as the choices of hybrid parameter $\lambda$. A distinctive feature of the auction design literature for autobidding auctions is the assumption that valuations are public instead of private as it is standard in the mechanism design literature. This assumption is predicated on the fact that advertisers increasingly rely on the machine learning algorithms that are developed by the advertising platforms to predict clicks and conversions.

\begin{table}[h]
    \centering
    \begin{tabular}{c|c|c|c|c}
      value $v_i$  & RoS target $\tau_i$ & budget $B_i$ & $\lambda$ & paper  \\
      \hline
      private & public & $B_i = \infty$ & $\lambda = 1$ & \citet{golrezaei2021auction}  \\
      \hline
      public & private & $B_i = \infty$ & $\lambda = 0$ or $\lambda = 1$ & \citet{balseiro2021landscape}  \\
      \hline
      private & public & $B_i = \infty$ & $\lambda = 0$ or $\lambda = 1$ & \citet{balseiro2021landscape}  \\
      \hline
      private & public & $B_i = \infty$ & $\lambda \in (0, 1)$ & ex-post RoS \citet{lv2023auction}  \\
      \hline
      private & private & $B_i = \infty$ & $\lambda = 0$ & deterministic \citet{balseiro2023optimal}  \\
      \hline
      private & public & public & $\lambda = 1$ & \citet{goel2014clinching} \\
      \hline
      public & private & public & $\lambda = 0$ &  \citet{balseiro2022optimal}  \\
      \hline
      public & private & private & $\lambda = 0$ & \citet{xing2023truthful}
    \end{tabular}
    \caption{Relevant works by the information structure on values, RoS target, and budget, as well as the bidding agent objective type (parameterized by $\lambda$).}
    \label{tab:auctions}
\end{table}

\subsubsection{RoS constraint only}
\citet{golrezaei2021auction} consider the revenue-optimal auction design for utility-maximization agents ($\lambda = 1$) with ROI constraints, which can be equivalently modeled with RoS constraints. They find empirically, some buyers in the online ad market behave as if they are subject to such constraints. In the symmetric setting where agents have the same RoS target, they show that an optimal auction is one of the following depending on the RoS target: (i) second-price auction with the Myersonian reserve price, (ii) second-price auction with a reduced reserve price, (iii) second-price auction without reserve plus a participation subsidy. In the general asymmetric case, the optimal auction is more complex and can be interpreted in terms of modified virtual values.

\citet{balseiro2021landscape} study the revenue-optimal mechanisms under different information structure on values and RoS targets for agents with either value-maximization objectives ($\lambda = 0$) or utility-maximization objectives ($\lambda = 1$). In the case of value-maximization ($\lambda = 0$), when either the values of agents are public information or the RoS targets of the agents are public information, they construct optimal mechanisms that achieve the first best (i.e., the optimal allocation when agent types are all public), which is not true in general when both values and RoS targets are private.
In contrast, for the case of utility-maximization ($\lambda = 1$), when either the values of agents are public information or the RoS targets of the agents are public information, they construct the corresponding optimal mechanisms, while the first best cannot be achieved.

\citet{lv2023auction} consider the revenue-optimal auction for bidding agents with intermediate objectives ($\lambda \in (0, 1)$) and require the RoS constraint to be satisfied ex-post instead of ex-ante, where the values of agents are private while the RoS targets of agents are public. They first provide a full characterization for dominant-strategy incentive compatibility: (i) monotone allocation rule and (ii) unique payment rule for any given monotone allocation. These can be seen as a generalization of Myerson's lemma \citep{myerson1981optimal}, while the unique payment rule follows a different relationship with the given allocation rule. They obtain the optimal auction for the single bidder case ($n = 1$) when a certain regularity condition is assumed (Decreasing Marginal Revenue).

\citet{balseiro2023optimal} prove that for the single value-maximization agent case ($n = 1$ and $\lambda = 0$), when both the values and RoS target are private information, the revenue-optimal mechanism with deterministic allocation can be implemented as a two-part tariff, i.e., a fixed price for buying the item and a fixed subsidy for not buying the item. An important implication from the structure of the optimal mechanism is that one does not need to screen the agent's RoS target.

\subsubsection{RoS and Budget constraints}

\citet{goel2014clinching} propose a generalized notion of admissible set that covers both the budget constraint and the RoS constraint. An admissible set can be modeled as $p_i \leq \alpha_i(x_i)$, where $\alpha_i$ is an increasing function. When it is a constant, it can capture the standard budget constraint, and when it is linear in $x_i$, it captures the RoS constraint. When it is the minimum of them, it captures both constraints at the same time. With the admissible set model, they consider the auction design with utility-maximization agents ($\lambda = 1$). In particular, they design a clinching auction \citep{ausubel2004efficient} that is incentive compatible, individually rational and Pareto-efficient.

\citet{balseiro2022optimal} study the case with the budget constraint in addition to the RoS constraint. They consider the case for value-maximization agents ($\lambda = 0$) where the values and budgets of the agents are public information while the RoS targets are private. They obtain the revenue-optimal mechanism for $n = 1$ and $n = 2$ in general cases, and the optimal mechanism for $n \geq 3$ for special cases.
Specifically, their optimal mechanism implements the efficient allocation according to RoS targets clipped up to thresholds depending on others' reports.

\citet{xing2023truthful} focus on the setting with value-maximization agents ($\lambda = 0$) where the values of agents are public information but both the RoS targets and budgets of the agents are private information. They provide the necessary and sufficient conditions for any allocation rule that can derive a truthful auction, and hence reduce the design space to allocation rules satisfying those conditions. Based on this characterization, they propose a family of simple truthful auctions. Although those auctions are not necessarily optimal, the characterization result is a non-trivial advancement towards this public value, private budget and RoS setting.

\subsection{Auction design with ML advice}

In online advertising, the auctioneer may have additional information about bidders' values via various machine learning technologies, i.e., ML advice. This additional information can be modeled as priors in a Bayesian setup as in Section \ref{subsec:bayesian}.

Alternatively, \citet{deng2021towards,balseiro2021robust,deng2022efficiency} take a prior-free approach and model this ML advice as an approximate signal $\in [\gamma \cdot \text{value}, \text{value})$ for each bidder. They show using this ML advice as reserves or boosts in VCG and FPA can significantly improve welfare efficiency (see Section \ref{subsubsec:reserve-boost} for more details). With ML advice as reserves, \citet{deng2024individual} demonstrate an individual welfare lower bound guarantee for this advertiser that increases in the advertiser's uniform bid multiplier, the quality of ML advice, and the relative market
share of this advertiser compared to competitors. Together with results in \citet{balseiro2021robust}, incorporating
ML advice as personalized reserves achieves “best of both worlds” by simultaneously benefiting total and individual welfare.

\subsection{Interdependent Auctions}

\citet{lu2023auction} consider a non-Bayesian model that is slightly different from our setting introduced in Section~\ref{sec:prelim}, where the allocation and payment in each single auction $j$ depend on the bids $\{\b_i\}_{i = 1}^n$ on all auctions. In other words, the allocation and payment of each single auction are no longer independent, instead, all the auctions are interdependent. They focus on constructing interdependent auctions with value-maximization agents ($\lambda = 0$) having a good competitive ratio compared against the offline optimal benchmark (i.e., no incentive constraints). They establish upper and/or lower bounds on the competitive ratios for several combinations across the information structure (fully private vs partially private), the demand type of agents (single-item, multi-item unit-demand, multi-item additive), and item divisibility.

\subsection{Auctions with Alternative RoS Constraint}

\citet{wilkens2016mechanism,wilkens2017gsp} initiate a line of work focusing on an alternative definition of the RoS constraint, where the constraint is enforced for each auction $j$ separately rather than the aggregation over all $m$ auctions. Formally, the alternative RoS constraint for each bidding agent $i$ is
\begin{equation}\label{eq:ros-alter}\tag{\textsc{RoS'}}
  x_{ij} \cdot v_{ij} \geq \tau_i \cdot p_{ij},\quad \forall j \in [m].
\end{equation}
Under this definition \eqref{eq:ros-alter}, \gsp is incentive compatible for the bidding agents, which in general is not the case with \eqref{prog:bidding-ros}.\footnote{\eqref{prog:bidding-ros} and \eqref{eq:ros-alter} are equivalent when there is only one auction ($m = 1$) and the constraint is ex-post.}

\citet{lv2023utility} consider the mechanism design problem with the alternative definition \eqref{eq:ros-alter} when agent with both utility-maximization and value-maximization objectives are present. When their objective types are public, they show that one can use the same efficient allocation rule (higher bids wins higher slots) for all agents and \vcg (\gsp) payment for utility-maximization (value-maximization) agents. When their objective types are private, they propose a novel mechanism such that the payment of each agent depends on its allocated slot but not their objective type. Under this mechanism, they also prove a $2$-approximation in terms of liquid welfare.

\section{Emerging Topics}\label{sec:emerging}

In this section, we cover some emerging topics in the literature that go beyond bidding algorithms, equilibrium and PoA analysis, and optimal auction design.

\subsection{Utility functions of advertisers using autobidding}

So far this survey has focused on the interaction between bidding agents and the platform (the auctioneer), assuming advertisers' inputs as fixed. However, to fully grasp the impact of auction formats, we must model how advertisers react. In game-theoretic language, most autobidding research has focused on the bidding agent subgame, neglecting the multi-period game where advertisers first submit inputs, followed by the subgame with the bidding agent decisions where the allocation and payment accrues.

The key question in modeling advertiser decisions is whether they are utility-maximizing, value-maximizing or something else. Auction design has traditionally assumed utility maximization, but the rise of target-based bidding strategies challenges this. Why would a utility maximizer use a value-maximizing bidding agent?  If instead advertisers' objective is to maximize value subject to a constraint, what incentives guide their input decisions to the autobidding agent?

Regarding the first question, one informal argument for value-maximization agents being favored in practice is a principal-agent model \citep{fadaei2017truthfulness,bichler2018principal}. In this model, each advertiser has a decision department (the principal) and an execution department (the agent) with slightly misaligned goals. To mitigate risk and ensure performance, the principal often sets value-maximization goals for the agent with clear constraints.

Recently, \cite{perlroth2023auctions} demonstrate that a utility-maximizing agent prefers to bid through a target-based bidding agent rather than through a marginal-based bidding agent when the platform {\em lacks commitment} to the declared auction rules: that is, the platform can revisit the rules of the auction (e.g., may readjust reserve prices depending on the bids submitted by
the bidders) after bids have already been submitted.
Furthermore, they show that due to the lack of commitment the bid shading effect when advertisers bid using a marginal bidding agent is so aggressive that if the platform would enforce to bid only through a marginal bidding agent (e.g. by removing the option of using a target-based autobidder), the platform's revenue would be lower than the revenue they obtain when advertisers use a target-based bidding agent.
\citet{bergemann2023} study the welfare and pricing implications when profit-maximizing advertisers use autobidding systems and lack user data which is known to the autobidder/platform. Compared to the case where advertisers can directly bid in each auction (and all user data is known to them), they show that the autobidding system create negative externalities on external advertising channels (outside of the platform) both in terms of allocation efficiency and consumer surplus.

If in turn advertiser's objectives are aligned with a value-maximizing objective,
\cite{alimohammadi2023incentive}
study what type of auctions are {\em autobidding incentive compatible (AIC)}: for what type of auctions an advertiser with a target-based preference (or a budget-based preference) prefers to submit their constraint as their input to the autobidding agent.
They show the second price auction is not AIC for both the target and budget case. For first-price auctions, when bidding agents are restricted to use a uniform policy the auction is AIC, while when they can also use non-uniform bidding strategies then auction is not AIC. More recently, \citet{feng2024strategic} investigate the PoA of running first-price auctions with budget-constrained autobidders when the budget constraints are strategically chosen by the advertisers and demonstrate constant PoA for such a game.

In addition, there is a second stream of literature on the multi-channel auction problem where they study how value-maximizing advertisers strategically submit their inputs to multiple autobidder agents, where each autobidder agent bids on advertiser's behalf for a particular channel. The following section presents the most interesting results on this topic.

\subsection{Multi-channel}

In practice, advertisers may procure ad impressions simultaneously on multiple advertising channels.
This can involve optimizing campaigns across a single platform's various channels (e.g., Google Ads inventory, including YouTube, Display, Search, Discover, Gmail, and Maps) or across channels owned by different platforms (such as Google, Meta, and Microsoft).
In such scenarios, if advertisers are value-maximizing agents subject to a
global \eqref{prog:bidding-ros} and \eqref{prog:bidding-budget} then the advertiser's bidding problem and the channel's auction design problem are far from trivial as the advertisers' global constraints interlinks the bidding problem (and, hence, the auction design) across channels.

In what follows, we present recent research that has been trying to shed light on this topic both from an advertiser perspective on how to bid across the channels as well as from a channel perspective on the design of auctions.

\subsubsection{Bidding with multiple channels}

\citet{deng2023multi} study the problem of multi-channel bidding where an advertiser aims to maximize their total conversion while satisfying aggregate \eqref{prog:bidding-ros} and \eqref{prog:bidding-budget} constraints across all channels. In particular, the advertiser can only utilize two levers on each channel to set up their campaigns, namely setting a per-channel budget and per-channel target RoS. \citet{deng2023multi} first analyze the effectiveness of each of these levers via comparison against the global optimum in which the advertiser can directly bid on each impression, and show that: when an advertiser only optimizes over per-channel RoSs, their total conversion can be arbitrarily worse than what they could have obtained in the global optimum, while the advertiser can achieve the global optimum leveraging per-channel budgets only. Under a bandit
feedback setting, \citet{deng2023multi} further present an efficient and low-regret learning algorithm that produces per-channel budgets whose resulting conversion approximates that of the global optimum. \citet{susan2023multi} present a strategy for multi-channel bidding when channels adopt auction rules that may or may not be incentive-compatible under the presence of budget constraints. \citet{aggarwal2024multi} characterize the optimal bidding for a continuous query-model where the size of a query is infinitesimal. They show that the advertiser's bidding problem is equivalent to finding a per channel uniform bid such that the advertiser's marginal cost-per-acquisition in each of the channels is the same.

\subsubsection{Multi-channel auction design}
\citet{aggarwal2023multi} initiate the study of multi-channel autobidding auction design focusing on the case of a platform owning multiple internal advertising channels (e.g., Google: Search, Play, YouTube; Meta: Instagram, Facebook, Messenger, etc.) In their setting, they allow a general advertising ecosystem with advertisers having either a \eqref{prog:bidding-ros} or \eqref{prog:bidding-budget} global constraint as well as profit-maximizing advertisers but restrict channels to sell their inventory using a SPA with a reserve price. They study the revenue implications for the platform of having each channel to independently optimize their reserve prices (local optimization) compared to having a global reserve price policy across the channels (global optimization).
They consider two models: one in which the channels have full freedom to set reserve prices, and another in which the channels have to respect floor prices set by the publisher. They show that in the first model, welfare and revenue loss from local optimization is bounded by a function of the advertisers' inputs, but is independent of the number of channels and bidders (see Theorem 3 on \citet{aggarwal2023multi} for details on the specific bounds).
For the second model, they show that the revenue from local optimization could be arbitrarily smaller than those from global optimization.

\citet{aggarwal2024multi} study the problem of auction design in the multi-channel setting where multiple platforms (each own a single channel) are competing to sell their inventory to the same pool of advertisers. They consider value-maximizing advertisers that have a \eqref{prog:bidding-ros} constraint across channels. The advertisers strategically report target ROIs to each channel's autobidder, which bids uniformly\footnote{{Note that, while uniform bidding is optimal when each channel is using a truthful auction \citet{aggarwal2019autobidding}, uniform bidding is generally not optimal when the channel is running a first-price auction.}
In this paper, uniform bidding is used to model a practical constraint that a system might impose.} on their behalf into the channel's auction. Each platform chooses between using a first-price auction or a second-price auction to maximize its own revenue. They show that for a revenue-maximizing platform, competition is a key factor to consider when designing auctions. While first-price auctions are optimal (for both revenue and welfare) in the absence of competition \citep{deng2021towards}, this no longer holds true in multi-channel scenarios. \citet{aggarwal2024multi} show that for the case of two competing platforms, there exists a large class of valuations for the advertisers such that from the platform's perspective,
running a second-price auction (rather than a first-price auction) is a dominant strategy.
They also identify some key factors influencing the platform's choice of auction format: (i) advertiser sensitivity to price changes -- how much the advertisers' reported targets change against auction changes, (ii) intensity of competition among advertisers, and (iii) relative inefficiency of second-price auctions compared to first-price auctions.

\subsection{Empirical Studies}

In the previous sections, we discussed the theoretical understanding of autobidding auctions in different aspects. However, the performance of different auction formats is usually analyzed in terms of PoA, which essentially focuses on welfare analysis in the worst-case scenarios, while the real-world instances could have much better equilibrium welfare. To complement the theoretical analysis, \citet{deng2024non} empirically study how different auction formats (namely VCG, FPA and GSP) perform in the autobidding world with synthetic datasets when advertisers adopt different bidding algorithms. 

\paragraph*{Non-uniform bid scaling}  \citet{aggarwal2019autobidding} demonstrate that uniform bid-scaling (i.e., always bid $\kappa v$ with a universal bid-scaling factor $\kappa$ when the bidder's value is $v$) is an optimal strategy for value maximizers in auctions that are truthful for quasi-linear utility maximizers. Therefore, each autobidding agent is only required to optimize one bid-scaling factor to find the best strategy. On the other hand, for auctions that are not truthful for quasi-linear utility maximizers (such as FPA and GSP), uniform bid-scaling can result in a suboptimal bidding strategy, while non-uniform bid-scaling (i.e., use different bid-scaling factors in different auctions) may greatly improve the bidding performance.

\paragraph*{Synthetic datasets and experiment setup} To generate the datasets that mimic the data structure of practical ad auctions, \citet{deng2024non} randomly draw query features and bidder features from multidimensional Gaussian distributions, and the bidder's values are drawn from log-normal distributions parameterized by query and bidder features. To facilitate the simulation of non-uniform bid-scaling algorithms, \citet{deng2024non} partition the queries to different categories following a multi-layer laminar structure. Each bidder chooses different bidding multipliers for different query clusters, and updates the multipliers through a gradient-descent based algorithm in each round. 

\paragraph*{Empirical Results} When bidders only adopt uniform bid-scaling strategies, it is observed that FPA $>$ GSP $>$ VCG for both welfare and profit. Such a result is consistent with the theoretical finding in the sense that FPA has better welfare and profit~\citep{deng2021towards}. 
When bidders can adopt non-uniform bid-scaling strategies, the empirical result of FPA $>$ GSP $>$ VCG for both welfare and profit continues to hold. For different levels of non-uniform bid-scaling algorithms, where a higher non-uniform bid-scaling level corresponds to a larger number of query clusters with different bid multipliers, there are different trends for different auction formats. For FPA, both profit and welfare decrease as the non-uniform bid-scaling level increases. 
On the other hand, for GSP, increasing the non-uniform bid-scaling level increases profit but decreases welfare; and for VCG, switching to different levels of non-uniform bid-scaling has no effect on welfare and profit.

\section{Conclusion}

In this survey, we covered a large portion of recent works related to autobidding in the online advertising ecosystem. We mentioned bidding algorithms for both truthful and non-truthful auctions in the presence of RoS and budget constraints. We discussed the existence of equilibrium, the price of anarchy with respect to different solution concepts, and the convergence properties of several bidding dynamics. We introduced recent advancements in terms of revenue-optimal auction design under different information structures and with various benchmarks. Finally, we discussed emerging topics in the literature, such as the role of advertiser decision, the application with multi-channel, and the comparison between theoretical and empirical results. We hope this survey provides a valuable resource for both practitioners and academics seeking to understand the state-of-the-art in this rapidly evolving field.

\bibliography{ref}
\bibliographystyle{apalike}

\end{document}